\documentclass[aps,showpacs,preprintnumbers,amsmath,amssymb]{revtex4}

\usepackage{graphicx}
\usepackage{dcolumn}
\usepackage{bm}
\input epsf.sty

\begin{document}
\title{Chiral Phase Transitions in QED at Finite Temperature: \\
Dyson-Schwinger Equation Analysis in the Real Time \\
Hard-Thermal-Loop Approximation}

\author{Yuko FUEKI}%
 \email{yfueki@phys.nara-wu.ac.jp}
 \affiliation{Department of Physics,  Nara Women's University,
   Kita Uoya-nishimachi, Nara 630-8506, Japan}%

\author{Hisao NAKKAGAWA}%
 \email{nakk@daibutsu.nara-u.ac.jp}
\author{Hiroshi YOKOTA}%
 \email{yokotah@daibutsu.nara-u.ac.jp}
\author{Koji  YOSHIDA}%
 \email{yoshidak@daibutsu.nara-u.ac.jp}
\affiliation{Institute for  Natural Sciences,  Nara University,
1500 Misasagi-cho,
 Nara 631-8502, Japan}

\date{\today}

\begin{abstract}
In order for clarifying what are the essential thermal effects that govern
the chiral phase transition at finite temperature, we investigate, in
 the real-time thermal QED, the consequences of the Hard-Thermal-Loop (HTL)
 resummed Dyson-Schwinger equation for the physical fermion mass function
 $\Sigma_R$.
Since $\Sigma_R$ is the mass function of an ``unstable''
quasi-particle in
thermal field theories, it necessarily has non-trivial imaginary parts
together with non-trivial wave function renormalization constants.
 The analyses so far completely
dismissed this fact, despite it being one of the basic outcome of
 thermal field theory. The ``approximation'', which neglects the
 imaginary parts of the $\Sigma_R$,
gives (non-trivial) constraint equations to be solved simultaneously,
 thus cannot
be consistently carried out except in the trivial case, $\Sigma_R =
 0$. In the
present analysis we  correctly respect this fact, and study, in the
 ladder
approximation, the effect of HTL resummed gauge boson propagator. Our
 results with
the use of numerical analysis, show the two facts;
i) The chiral phase transition
is of second order,  since the fermion mass is dynamically generated
at a critical
value of the temperature $T_c$,  or at the critical coupling constant
$\alpha_c$,
without any discontinuity, and ii) the critical temperature $T_c$ at
fixed value of
$\alpha$ is significantly lower than the previous results, namely the
restoration
of chiral symmetry occurs at lower temperature than previously
expected.
The second fact shows the importance of correctly taking the essential
thermal
effect into the analysis of chiral phase transition, which are, in the
previous
analyses, neglected due to the inappropriate approximations.  The
procedure how
to maximally respect the gauge invariance in the present
approximation, is also
discussed.
\end{abstract}

\maketitle

\makeatletter 
\@addtoreset{equation}{section}
\def\theequation{\thesection.\arabic{equation}}

\section{Introduction}

Although lots of efforts have been made to undestand the temperature- and/or
density-dependent phase transition in thermal QCD/QED,  we cannot have yet truly
understood even the relation between the chiral transition and the
confinement-deconfinement transition.  Beginning of the relativistic heavy ion
collision experiments at BNL-RHIC, which has been actively producing new results
that may prove the generation of the quark-gluon plasma phase, has attracted an
increasing interest in studying the physics in thermal QCD, thus has given us an
encouraging time to proceed to further investigations of the mechanism of phase
transition in hot and dense gauge theories, especially in QCD and QED.

The Dyson-Schwinger (DS) equation is proven to be a powerful tool to investigate
\textit{with the analytic procedure} the phase structure of gauge theories,
especially in the vacuum gauge theories [1,2,3]. However, we cannot say that,
at finite temperature and/or density, the DS equation analyses of chiral and/or
di-quark condensation have been carried out successfully.  

In the preceding DS equation analyses [4-8], the lessons from  vacuum theories
have been so simply applied to thermal theories without close examination. In most
analyses the ladder approximation was used by simply neglecting all the
hard-thermal-loop (HTL) effects [5,7], or only by taking  improperly the HTL
effects into the gauge boson propagator [6,8]. As a result they have missed the
essential contribution of thermal gauge field theories, especially the important
effect from the ``dymanically screened'' magnetic mode (having in general a
momentum-dependent ``mass'', though being massless in the static limit).  Many
analyses, by fixing in the Landau gauge, ignored the fermion wave-function
renormalization constants (WFRCs) by taking their naive tree values [5,6,7].
Furthermore, most analyses done in the real time formalism did not discuss the
physical fermion mass function $\Sigma_R$ itself of the retarded propagator [6,8],
without any respect to the existence of imaginary parts,
together with the inaccurate use of
the instantaneous exchange (IE) approximation to the gauge boson propagation [6,8].
All such improper approximation methods may have caused the neglection of
would-be-large contributions to the DS equation otherwise existed.
   
Then we should seriously ask whether we could rely on the previous results of the
DS equation analysis on the chiral phase transition as the real consequences of
thermal gauge field theories. Considering the troubles in the previous analyses [4-8]
mentioned above, we should make a re-analysis by studying the HTL
resummed DS equation in the real time formalism, thus might giving a new
understanding on the phase structure and the mechanism of phase transition in
thermal gauge theories.

Main interest of the present investigation lies in clarifying what are the
essential temperature effects that govern the phase transition and also in
finding how we can closely take these  effects into the ``kernel'' of the DS
equation. Essential procedures of our analysis can be summerised as follows; 

i) Firstly we use the real-time closed-time-path (RTCTP) formalism [9], and study
the physical mass function $\Sigma_R$ itself, not the $\Sigma_{11}$, of the retarded
fermion propagator. 

ii) Secondly we correctly respect the fact that the $\Sigma_R$ is the mass function
of an ``unstable'' quasi-particle in thermal field theories, namely that the $\Sigma_R$
has non-trivial imaginary parts as well as non-trivial WFRCs. Neglection of imaginary
parts and non-trivial WFRCs actually give (non-trivial) constraint equations to be
solved simultaneously. This fact was totally dismissed in the preceding analyses.

iii) Thirdly  and most importantly, devoting our attension to closely estimating
the dominant temperature-dependent contributions, we focus on studying the DS
equation being exact up to HTL approximation: Both the gauge boson propagators
and the vertex functions are determined within the HTL resummation [10,11,12,13],
with which the gauge invariance of the result at least in the perturbative analysis
is guaranteed.  With the HTL resummed vertex functions we can explicitly write down
the HTL resummed DS equation [12]. 

iv) Finally in connection with ii) and iii) above, the gauge-paremeter dependent
contribution must be carefully studied without fixing the gauge into some definite
ones, such as the Landau gauge.   

The third point listed above is better to be taken step by step into the actual
analysis of the DS equation.  The advantage of the DS equation analysis lies in
the possibility of systematic (step-by-step) improvement of the approximation to
the integration kernel through the analytic investigation. 

Thus, in the present paper we present the result of our first step investigation in
the strong coupling QED; focussing on what happens when we take into account exactly
the HTL resummed gauge boson propagators, we investigate the consequences of the ladder
(point vertex) DS equation with the use of the fully HTL resummed photon propagator.
In this approximation, however, the gauge invariance of the results is spoiled. The
procedure how to maximally respect the  gauge invariance of the results in the ladder
approximation, is discussed and the result is given in a separate paper[14]. Analysis
in QCD and effects of fully including the HTL resummed vertices will also be presented
in the separate paper [15]. 

This paper is organized as follows. In the next section II, the DS equation in QED for
the fermion self-energy in the HTL approximation is derived, and the improved ladder
(point vertex) DS equation with the HTL resummed photon propagator is given. The
instantaneous exchange (IE) approximation for the photon propagation is disccused.
In section III we give the results of the present investigation with the use of
numerical analysis: the order of the chiral phase transition, the critical curve in
the $T$-$\alpha$ plane (where $T$ is the temperature of the environment and $\alpha$
the coupling constant), and values of the critical exponents. Conclusion of the
present analysis and discussion on the results are given in the last section IV.

\section{The Dyson-Schwinger equation in QED for the fermion self-energy function}

\subsection{The DS equation in the Hard-Thermal-Loop approximation} 

The DS equation for the physical, i.e., the retarded fermion self-energy function
$\Sigma_R$ in the HTL approximation can be obtained by applying the following
approximation to the full DS equation;

   i) replace the full gauge boson propagator with the HTL resummed propagator, and

   ii) approximate the full vertex functions to the HTL resummed vertex functions.

Then in the RTCTP formalism we get in QED the desired DS equation (in QCD, the
running coupling should be used inside the loop-momentum integration) [12],

\begin{eqnarray}
\label{eq_1} 
& & - i\Sigma_R (P)  = -i \Sigma_{RA}(-P,P) =
        - \frac{e^2}{2} \int \frac{d^4K}{(2\pi)^4} 
       \nonumber \\
  &  & \ \ \ \times \left[ ^*\Gamma^{\mu}_{RAA}(-P,K,P-K) S_{RA}(K) 
       \ ^*\Gamma^{\nu}_{RAA}(-K,P,K-P) \ ^*G_{RR,\mu\nu} (P-K) 
       \right. \nonumber \\
  & & \ \ \ \ \left. + \ ^*\Gamma^{\mu}_{RAA}(-P,K,P-K) S_{RR}(K)
       \ ^*\Gamma^{\nu}_{AAR}(-K,P,K-P) \ ^*G_{RA, \mu\nu}(P-K)
       \right]\ ,
\end{eqnarray}
where $^{\ast}G^{\rho \sigma}$ is the HTL resummed gauge boson propagator, whose
retarded $R \equiv RA$ component is given by 
\begin{equation}
\label{eq_2}
  \ ^{*}G^{\rho \sigma}_R(K)= \frac{1}{\ ^{*}\Pi^R_T(K)-K^2-i \epsilon k_0}
     A^{\rho \sigma} + \frac{1}{\ ^{*}\Pi^R_L(K)-K^2-i \epsilon k_0}
     B^{\rho \sigma} - \frac{\xi}{K^2+i \epsilon k_0}
     D^{\rho \sigma} \ ,
\end{equation}
and the $RR$ component by
\begin{equation}
\label{eq_3}
 \ ^{*}G^{\rho \sigma}_{RR}(-K,K)
         = (1+2n_B(k_0))[\ ^{*}G^{\rho \sigma}_R(K)
                                            - \ ^{*}G^{\rho \sigma}_A(K)], \ 
  n_B(k_0)= \frac{1}{\exp(k_0/T) - 1},
\end{equation}
with $\ ^{*}\Pi^R_T$ and $\ ^{*}\Pi^R_L$ being the HTL contributions to the
retarded gauge boson self-energy of the transverse and longitudinal modes [16],
respectively.  $ A^{\rho \sigma}$, $B^{\rho \sigma}$ and $D^{\rho \sigma}$ are
the projection tensors [17], 
$$
 A^{\rho \sigma}=g^{\rho \sigma} - B^{\rho \sigma}- D^{\rho \sigma}, \ 
 B^{\rho \sigma}=- \frac{\tilde{K}^{\rho} \tilde{K}^{\sigma}}{K^2} , \ 
 D^{\rho \sigma}= \frac{K^{\rho} K^{\sigma}}{K^2},  
$$
where $\tilde{K}=(k, k_0{\bf \hat{k}})$, $k=\sqrt{{\bf k}^2}$
and ${\bf \hat{k}}={\bf k}/k$ denotes the unit three vector along  ${\bf k}$.

$S(-P,P)\equiv S(P)$ denotes the full fermion propagator, whose retarded $RA
\equiv R$ component is given by
\begin{equation}
\label{eq_4}
 S_R(P)=\frac1{P\!\!\!\!/ + i\epsilon \gamma_0 - \Sigma_ R},
\end{equation}
and $S_{RR}$ the $RR$ component
\begin{equation}
\label{eq_5}
 S_{RR}(P)=(1-2n_F(p_0))[S_R(P)-S_A(P)], \
 n_F(p_0)= \frac{1}{\exp(p_0/T) + 1}.
\end{equation}
The fermion self-energy function $\Sigma_R$ in Eq.(\ref{eq_4}) can be tensor-decomposed
as
\begin{equation}
\label{eq_6}
  \Sigma_R(P)=(1-A(P))p_i\gamma^i -B(P)\gamma^0+C(P) \ ,
\end{equation}
with  $A(P)$, $B(P)$ and $C(P)$ being the three independent scalar invariants
to be determined.  At zero temperature, the wave function renormalization
constant $A(P)$ coincides with $B(P)$ and  equals to unity in the Landau gauge,
while at finite temperature it is not. $C(P)/A(P)$ plays the role of mass function,
in which we are interested.

$\ ^*\Gamma^{\mu}$ is the HTL resummed 3-point fermion-gauge boson vertex
function [12],
\begin{equation}
\label{eq_7}
  \ ^*\Gamma^{\mu}_{\alpha \beta \gamma}
       =\gamma^{\mu}_{\alpha \beta \gamma}
         +\delta \Gamma^{\mu}_{\alpha \beta \gamma},  \ \ 
        (\alpha, \beta, \gamma =A,R),
\end{equation}
where $\delta \Gamma^{\mu}_{\alpha \beta \gamma}$ represents the HTL
contribution to the 3-point fermion-gauge boson vertex function,
and $\gamma^{\mu}_{\alpha \beta \gamma}$ the tree vertex with
$\gamma^{\mu}_{RAA}=\gamma^{\mu}_{AAR}=\gamma^{\mu}$, otherwise zero.
Appearence of the HTL resummed vertex functions together with the HTL resummed
gauge boson propagators assures that the HTL approximation is consistently
carried out in studying the HTL resummed DS equation [12], and guarantees the
result being gauge invariant, at least, in the effective perturbation regime. 

\subsection{The improved ladder DS equation with the HTL resummed gauge boson
propagator}

Neglection of $\delta\Gamma^{\mu}_{\alpha \beta \gamma}$ in Eq.~(\ref{eq_7}),
simply brings us to the ladder (point-vertex) DS equation with the HTL resummed
gauge boson propagator.  It significantly simplifies the structure of the DS
equation to be examined, thus reducing the technical  difficulty in handling the
DS equation itself.  The price to pay is to lose the assurance of  gauge invariance
of the results. 

In this paper  as already mentioned before, we investigate the consequences of
the ladder DS equation with the HTL resummed gauge boson propagator, Eqs.~(\ref{eq_2}) and
(\ref{eq_3}).  In this case three invariants $ A(P)$, $B(P)$ and $C(P)$ as functions of
$p_0$ and ${\bf p}$ satisfy the following coupled integral equations
($P^{\mu}=(p_0, {\bf p})$),   

\newpage
\begin{eqnarray}
\label{eq_8}
  -p^2[1-A(P)] &=& -e^2 \left. \int \frac{d^4K}{(2 \pi)^4}
       \right[ \{1+2n_B(p_0-k_0) \} Im[\ ^*G^{\rho \sigma}_R(P-K)]
       \times  \nonumber \\
  & & \Bigl[ \{ K_{\sigma}P_{\rho} + K_{\rho} P_{\sigma}
       - p_0 (K_{\sigma} g_{\rho 0} + K_{\rho} g_{\sigma 0} ) 
       - k_0 (P_{\sigma} g_{\rho 0} + P_{\rho} g_{\sigma 0} )
       + pkz g_{\sigma \rho} \nonumber \\
  & & + 2p_0k_0g_{\sigma 0}g_{\rho 0} \}\frac{A(K)}{[k_0+B(K)+i
       \epsilon]^2 - A(K)^2k^2 -C(K)^2 }
       + \{ P_{\sigma} g_{\rho 0} + P_{\rho} g_{\sigma 0} \nonumber \\
  & & - 2p_0 g_{\sigma 0} g_{\rho 0} \}
       \frac{k_0+B(K)}{[k_0+B(K)+i \epsilon]^2 - A(K)^2k^2
       -C(K)^2 } \Bigr] + \{1-2n_F(k_0) \}
       \times \nonumber \\ 
  & & \ ^*G^{\rho \sigma}_R(P-K) Im \Bigl[
       \{ K_{\sigma}P_{\rho}  + K_{\rho} P_{\sigma} - p_0 (K_{\sigma}
       g_{\rho 0} + K_{\rho} g_{\sigma 0} ) - k_0 (P_{\sigma}
       g_{\rho 0} + P_{\rho} g_{\sigma 0} ) \nonumber \\
  & & + pkz g_{\sigma \rho} + 2p_0k_0g_{\sigma 0}g_{\rho 0}\}
       \frac{A(K)}{[k_0+B(K)+i \epsilon]^2 - A(K)^2k^2-C(K)^2 } 
       \nonumber \\
  & & \left. +  \{ P_{\sigma} g_{\rho 0} + P_{\rho} g_{\sigma 0}
       - 2p_0 g_{\sigma 0} g_{\rho 0} \}
       \frac{k_0+B(K)}{[k_0+B(K)+i \epsilon]^2 - A(K)^2k^2
       -C(K)^2 } \Bigr] \right] \ , \\
\label{eq_9}
  - B(P) &=& -e^2 \left. \int \frac{d^4K}{(2 \pi)^4} \right[
        \{1+2n_B(p_0-k_0)\} Im[\ ^*G^{\rho \sigma}_R(P-K)] \times
         \nonumber \\
  & & \Bigl[ \{ K_{\sigma} g_{\rho 0} + K_{\rho} g_{\sigma 0}
       - 2k_0 g_{\sigma 0} g_{\rho 0} \}
       \frac{A(K)}{[k_0+B(K)+i \epsilon]^2 - A(K)^2k^2
       -C(K)^2 } \nonumber \\
  & & + \{ 2g_{\rho 0} 2g_{\sigma 0} - g_{\sigma \rho} \} 
       \frac{k_0+B(K)}{[k_0+B(K)+i \epsilon]^2 - A(K)^2k^2-C(K)^2 }
       \Bigr] + \{1-2n_F(k_0) \} \times \nonumber \\ 
  & & \ ^*G^{\rho \sigma}_R(P-K) Im \Bigl[ \frac{A(K)}{[k_0+B(K)+i
       \epsilon]^2 - A(K)^2k^2 -C(K)^2 } 
       \{ K_{\sigma} g_{\rho 0} + K_{\rho} g_{\sigma 0}  \nonumber \\
  & & \left. - 2k_0 g_{\sigma 0} g_{\rho 0} \} + \frac{k_0+B(K)}{[k_0+B(K)+
       i \epsilon]^2 - A(K)^2k^2-C(K)^2 }
       \{ 2g_{\rho 0} 2g_{\sigma 0} - g_{\sigma \rho} \} \Bigr] 
       \right] \ , \\
\label{eq_10}
  C(P) &=& -e^2 \int \frac{d^4K}{(2 \pi)^4} g_{\sigma \rho} 
       \{1+2n_B(p_0-k_0) \} Im[\ ^*G^{\rho \sigma}_R(P-K)]
       \times \nonumber \\
  & & \Bigl[ \frac{C(K)}{[k_0+B(K)+i \epsilon]^2 - A(K)^2k^2
       -C(K)^2 } + \{1-2n_F(k_0) \} \times \nonumber \\ 
  & & \left. \ ^*G^{\rho \sigma}_R(P-K) Im \Bigl[
       \frac{C(K)}{[k_0+B(K)+i \epsilon]^2 - A(K)^2k^2
       -C(K)^2 } \Bigr] \right] \ ,
\end{eqnarray}
where the cutoff scale $\Lambda$ is introduced in order to regularize the integrals over
$k_0$ and ${\bf k}$.

\subsection{The ladder DS equation in the improved instantaneous exchange (IE)
apporoximation}

The DS equations, Eqs.~(\ref{eq_8})-(\ref{eq_10}), are still quite tough to be attacked, forcing us
further approximations for the analysis to be effectively carried out. However,
the approximation made use of must be consistent with the HTL approximation, without
missing the important thermal effects out of the kernel of the DS equation. 

Here it is worth noticing that the instantaneous exchange (IE) approximation to the
gauge boson propagation, frequently used in the preceding analyses [6,7,8], is
\textit{not compatible} with the HTL approximation in a strict sense. In the
exact IE-limit the HTL resummed transeverse mass function, $\ ^{\ast}\Pi^R_T(K)$,
vanishes and the transeverse (magnetic) mode becomes totally massless. Namely the
IE approximation discards the important thermal effect coming from the Landau
damping, thus dismissing the dynamical screening of the magnetic mode. This causes
the famous quadratic divergence of the Rutherford scattering cross section. 

To see this point more clearly, let us take the IE-limit of the DS equation,
Eqs.~(\ref{eq_8})-(\ref{eq_10}), and neglect $Im[A(P)]$ and $Im[C(P)]$, then we obtain the following
equation for $Im[B(P)]$,
\begin{eqnarray}
\label{eq_11}
 Im[B(P)] &=& \frac{e^2}{4 \pi} m_g^2 T \int_0^{\infty} k^2 dk \int_{-1}^1
  dz \frac{1}{E} \left( \frac{1}{[E^2+m_g^2]^2} + \frac{1}{E^4} \right)
  , \\
 m_g^2 & \equiv & \frac13 e^2T^2 \ , \ \ \ E \equiv \sqrt{ ({\bf p}-{\bf k})^2 } \ , \nonumber
\end{eqnarray} 
showing $Im[B(P)]$ to be quadratically divergent. In order to ignore the $Im[B(P)]$,
we must disregard the divergence and set $\infty = 0$. This fact means that we
cannot naively neglect the imaginary parts of the $\Sigma_R$ without facing the
inconsistent constraint equations.  

The reason why in the previous analyses this divergence did not appear, is that
the imaginary part of $\Sigma_R$ was completely neglected there from the beginning,
namely that the equation for $Im[\Sigma_R]$ was totally disregarded. As is evident
from the above, however, the ``approximation'' made use of in the previous analyses
[6,8] is not the consistent approximation. 

Taking the above into account, the approximation we further make use of is the
improved IE approximation to the longitudinal gauge boson propagator, by keeping
the exact HTL resummed transeverse propagator. In the IE-limit the HTL resummed
longitudinal mass function, $\ ^{\ast}\Pi^R_L(K)$, has a definite thermal mass
$m_g^2 \sim (eT)^2$, representing the Debye screening due to thermal fluctuation,
thus even in the IE limit the longitudinal mode can take into acount the essential
thermal effect. In the present analysis the gauge is fixed to the Landau gauge
($\xi =0$).

Here it is fair to note that in the point vertex ladder approximation, as already
mentioned before, the gauge invariance of the results is spoiled. To maximally
respect the gauge invariance, we should solve Eqs.~(\ref{eq_8})-(\ref{eq_10}) with the constraint
$A(P)=1$, which guarantees $Z_2=1$, being consistent with the Ward identity
$Z_1=Z_2$. This can be done by successively adjusting the gauge-parameter $\xi$
in solving the above equations (\ref{eq_8})-(\ref{eq_10}). The result of this gauge-parameter
dependent analysis will be published in a separate paper [14].

\section{Results of the numerical analysis}
\subsection{Numerical solutions}

Now we should solve numerically the DS equations Eqs.~(\ref{eq_8})-(\ref{eq_10}) with the IE
approximation to the longitudinal mode [18]. As we have already stressed, we
respect the existence of the non-trivial imaginary parts, $Im[A]$, $Im[B]$ and $Im[C]$,
together with the corresponding real parts, $Re[A]$, $Re[B]$ and $Re[C]$. We use the
numerical method consisting of starting with suitable trial functions for the
solution and iterating the calculational procedure until stable solutions are
obtained. This method is simpler to handle and is useful so long as the convergence
of the iteration is guaranteed. 

We choose several types of the trial functions: various choices of constants,
independent of $p_0$ and $p$.

At each iteration, the three-fold integration is performed, namely over $k$, over
$k_0$, which are cut off at the mass scale $\Lambda$, and over $z=\cos \theta$. The
integration kernel of the present DS equation shows a little bit singular behavior,
and the numerical integration of such a singular integrand needs careful integration
prescription, which is properly managed. As a result, in each calculation we
performed at most 1000 times iterations, and obtained fairly stable solutoins. 

Result of the present analysis shows that the wave function renormalization
constants receive 10-20 percent corrections, see Figure~1 and Table~I, indicating
the necessity of gauge-parameter dependent analysis [14]. The generated size of
the imaginary part is nearly the same as that of the real part, showing the
existence of the non-trivial imaginary parts, which is as expected though
completely neglected in the previous analyses. 
\begin{figure}[htbp]
\begin{center}
\epsfxsize=12cm
\epsfbox{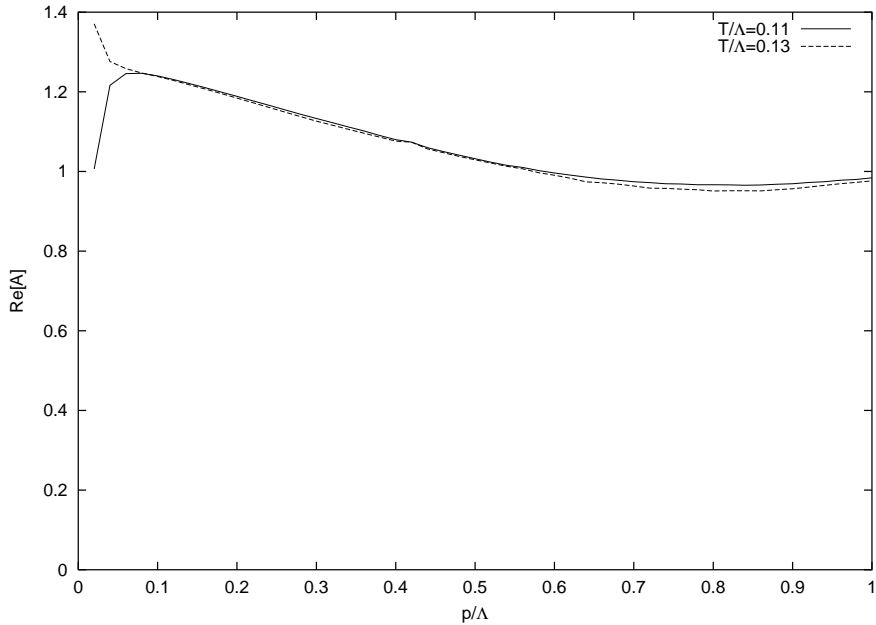}
\caption{Typical shape of the function $Re[A(P)]$ at $p_0=0$ for
                $\alpha=4.0$.}\label{fig_1}
\end{center}
\end{figure}

\begin{table}
\caption{Sample values of invariants, $Re[A(P)]$, ........., $Im[C(P)]$, 
      at $p_0=0$, $p=0.1\Lambda$ for $\alpha=4.0$.}\label{tab_1}
\begin{center}
\begin{tabular}{crrrrrr}
\hline
 $T$ & $Re[A(P)]$ & $Im[A(P)]$ & $Re[B(P)]$ & 
              $Im[B(P)]$ & $Re[C(P)]$ & $Im[C(P)]$ \\
\hline \hline
0.130 & 1.262 & 0.125 & -0.547 & 0.303 & 0.496 & -0.035 \\
0.135 & 1.227 & 0.067 & -0.551 & 0.311 & 0.466 & -0.057 \\
0.140 & 1.248 & 0.008 & 0.496 & 0.347 & -0.341& -0.071 \\
0.145 & 1.249 & 0.016 & 0.416 & 0.342 & -0.157 & -0.042 \\
0.150 & 1.217 & 0.059 & 0.378 & 0.315 & 0.000 & 0.000 \\
\hline
\end{tabular}
\end{center}
\end{table}

\subsection{Chiral phase transitions}

Let us now show the behavior of the mass function $M(P) \equiv Re[C(P)/A(P)]$
at some fixed value of $p$ as a function of the parameters $\alpha=e^2/(4\pi)$
and $T$. All the data shown in this section are those calculated at $p_0=0$,
which is suitable for studying the static mass.
The $T$-dependence of the mass function $M(P) \equiv Re[C(P)/A(P)]$ with
$p=0.1\Lambda$ is shown in Figure 2 for various fixed values of $\alpha$,
and in Figure 3 the $\alpha$-dependence of the mass function $M(P) \equiv
Re[C(P)/A(P)]$ with $p=0.1\Lambda$ is shown for various fixed values of $T$.
The errors resulting from the fluctuations are smaller than the size of the
symbol used for each sample point in Figures 2 and 3. 
\begin{figure}[htbp]
\begin{center}
\epsfxsize=12cm
\epsfbox{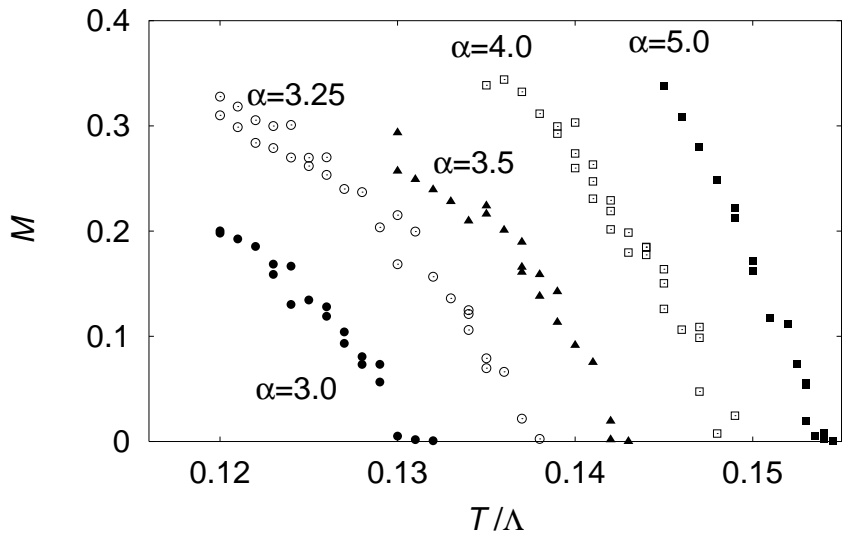}
\caption{The $T$-dependence of the mass function  $M(P) \equiv    
                Re[C(P)/A(P)]$  with $p=0.1\Lambda$  for various fixed values of 
                coupling constant $\alpha$.}\label{fig_2}
\end{center}
\end{figure}
\begin{figure}[htbp]
\begin{center}
\epsfxsize=12cm
\epsfbox{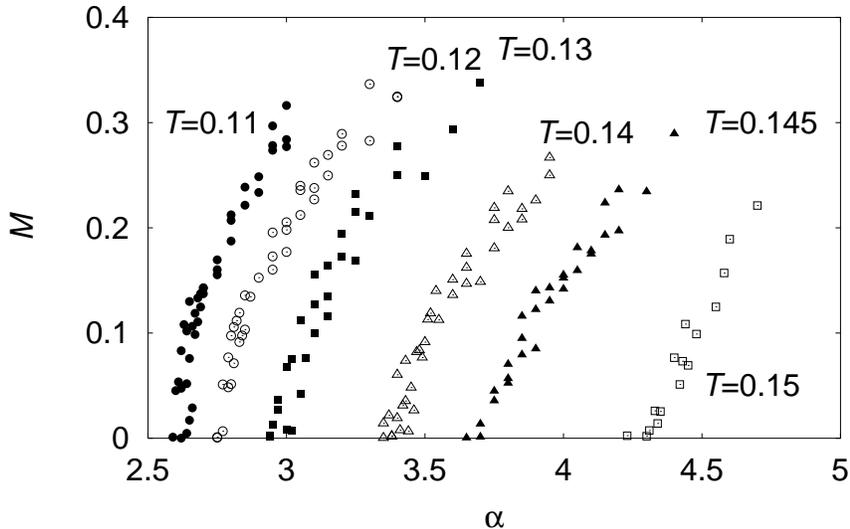}
\end{center}
\caption{The $\alpha$-dependence of the mass function  $M(P) \equiv 
                Re[C(P)/A(P)]$ with $p=0.1\Lambda$ for various fixed values of 
                temperature $T$.}\label{fig_3}
\end{figure}

From these figures we can see the two facts; i) The chiral phase transition is of
second order,  since a fermion mass is generated at a critical value of the
temperature $T$ or at the critical coupling constant $\alpha$ without any
discontinuity, and ii) the critical temperature $T_c$ at fixed value of $\alpha$
is significantly lower than the previous results [6,7,8], namely the restoration
of chiral symmetry occurs at lower temperature than previously expected. The
second fact shows the importance of correctly taking the essential thermal effect
into the analysis, which was disregarded in 
the previous analyses due to the inappropriate approximations.

\subsection{Critical curve and critical exponents}

Analyzing those data, typically shown in Figures 2 and 3 in the last subsection,
we can draw the critical curve that shows the phase-boundary in the
$(T,\alpha)$-plane. The critical curve and the critical exponents can be
determined systematically from the data with the use of the following method.

We fit the mass function $M(P)$ with $n$ data points, $M_i$, $\alpha_i$ and
$T_i$, with $i=1,2, .... n$ near the critical point by assuming the functional
form
\begin{equation}
\label{eq_12}
 M(P) = e^{C_T}(T_c - T)^{\nu} \ ,
\end{equation}
for fixed coupling constant $\alpha$, and
\begin{equation}
\label{eq_13}
 M(P) = e^{C_{\alpha}}(\alpha - \alpha_c)^{\eta} \ ,
\end{equation} 
for fixed temperature $T$. Here, $C_T$, $C_{\alpha}$, $T_c$, $\alpha_c$, $\nu$ and
$\eta$ are adjustable parameters, with $T_c$ and $\alpha_c$ corresponding to
the critical temperature and critical coupling constant, and $\nu$ and $\eta$
designate the critical exponents. The values $C_T$, $C_{\alpha}$, $T_c$,
$\alpha_c$, $\nu$ and $\eta$ can be determined with the method of the
standard least-squares fit.

\begin{figure}[htbp]
\begin{center}
\epsfxsize=12cm
\epsfbox{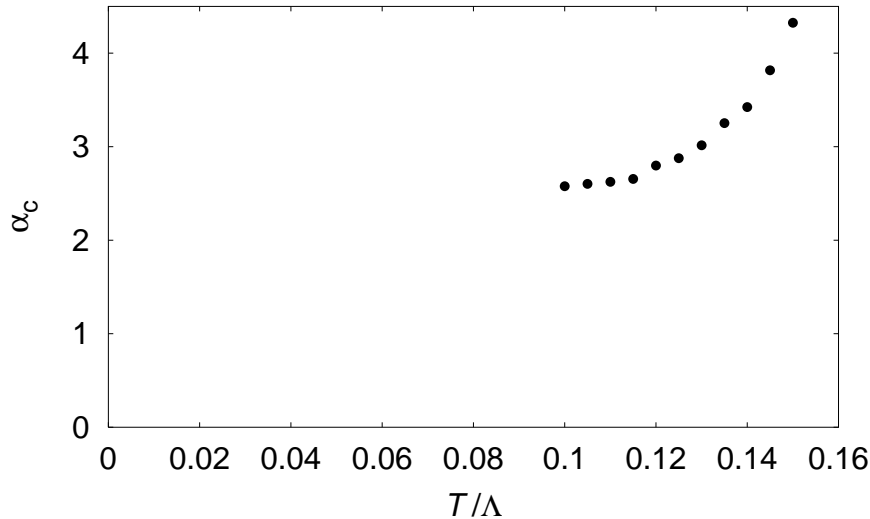}
\caption{The critical coupling constant $\alpha_c$ as a function of $T$.}
\label{fig_4}
\end{center}
\end{figure}
\begin{figure}[htbp]
\begin{center}
\epsfxsize=12cm
\epsfbox{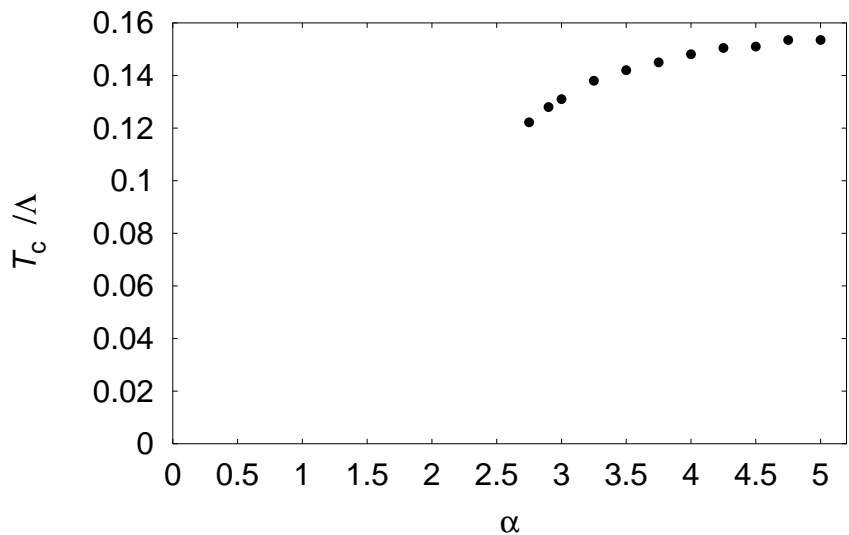}
\end{center}
\caption{The critical temperature $T_c$ as a function of $\alpha$.}
\label{fig_5}
\end{figure}
\begin{figure}[htbp]
\begin{center}
\epsfxsize=12cm
\epsfbox{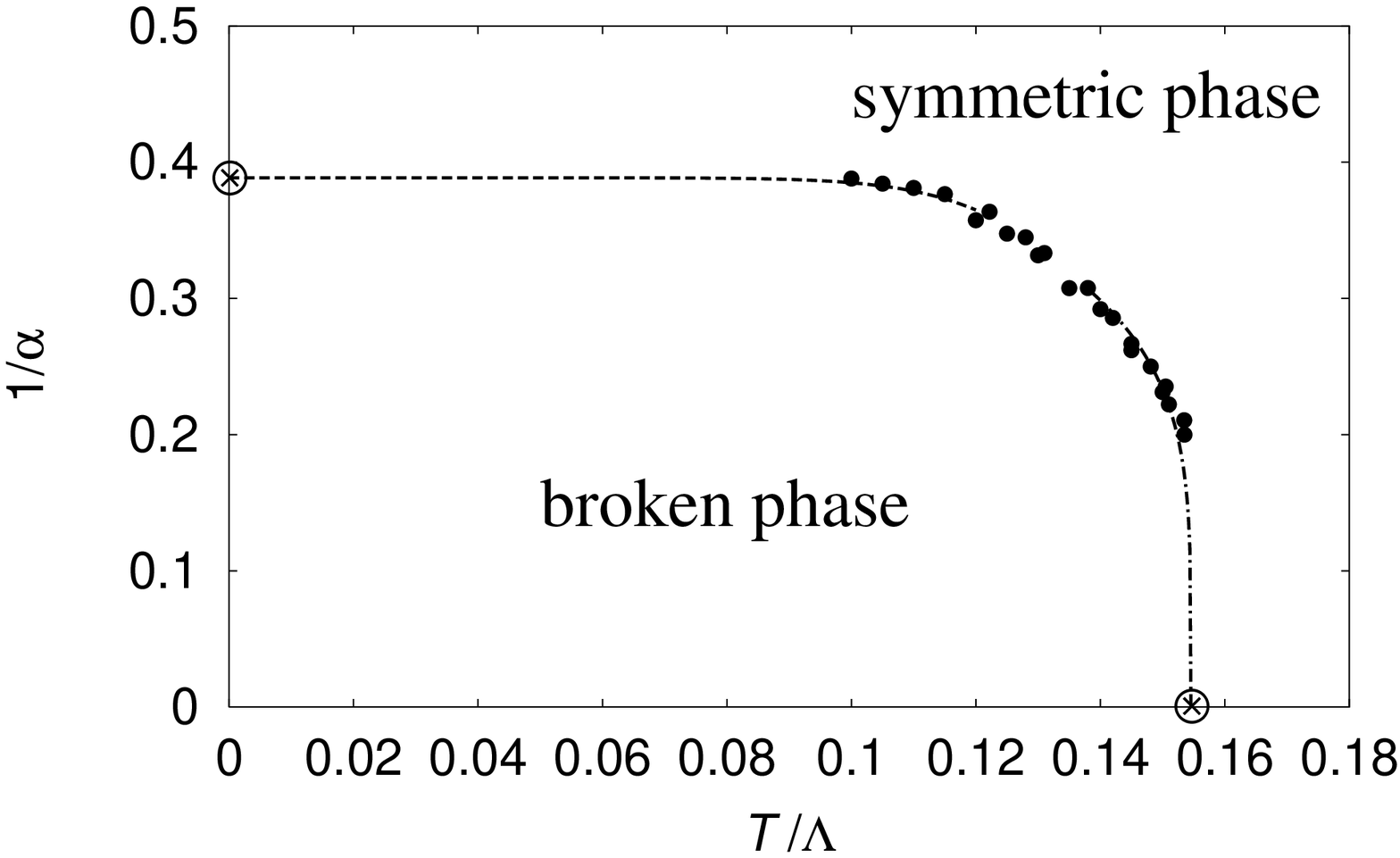}
\end{center}
\caption{The critical curve in the $(T,\alpha^{-1})$-plane. The dashed curve shows the fit of the critical curve for $T \to 0$, and the dash-dotted
one the fit for $\alpha \to \infty$ $(1/\alpha \to 0)$. The estimated
 values, $\alpha_c(T\to 0)$ and $T_c(1/\alpha \to 0)$ are shown with the
 symbol $\bigotimes$, see text.}
\label{fig_6}
\end{figure}

The critical coupling constant $\alpha_c$ as a function of $T$, and the critical
temperature $T_c$ as a function of $\alpha$, thus determined are shown in
Figures~4 and 5.  By putting them together we can determine the critical curve
in the $(T,\alpha^{-1})$-plane, Figure~6,
which shows two characteristic behaviors: a) As $T$ becomes smaller, the critical
coupling constant $\alpha_c$ also becomes smaller and seems to consistently
decrease from above to the zero temperature result, and b) the critical temperature
$T_c$ increases as a fuction of the coupling constant $\alpha$, with the possible
saturating behavior in the strong coupling limit, namely the critical curve suggests
the existence of the critical temperature, above which the chiral symmetry is always
restored. 

With the data in the the critical curve we may estimate the critical coupling constant
$\alpha_c$ in the limit $T \rightarrow 0$, $\alpha_c(T \rightarrow 0)$, by assuming
the functional form $(\alpha - \alpha_c(T \rightarrow 0)) \propto
T^{\beta}$. 
The fitting curve for $T \to 0$ is shown in Figure~6, by the dashed curve,
with
the estimated value $\alpha_c(T \to 0) = 2.58$ $ (1/\alpha_c(T \to 0) =
0.388)$.
The
estimated value, $\alpha_c(T \rightarrow 0) = 2.58$, is, however, significantly
larger than the value $\alpha_c(T=0)=\pi/3$ determined by theoretical analyses [1] of
the DS equation for the fermion self-energy part $\Sigma(P)$ at zero temperature, $T=0$,
in the ladder approximation in the Landau gauge with the tree level photon propagator. 

This result could be expected because of the approximation we have made use of in the
present analysis. As we have already noticed, even if the photon propagator is HTL
resummed, we adopted the improved IE approximation to the photon propagation. This
approximation actually limits the applicability of our analysis to the rather high
temperature region. In fact, at lower temperatures $T < 0.1\Lambda$, the stability
of the solution is getting spoiled.

We may also estimate the value of the critical temperature $T_c$ in the strong
coupling limit $\alpha \rightarrow  \infty$, $T_c(\alpha \rightarrow  \infty)$, by
assuming the functional form $(T_c(\alpha \rightarrow  \infty) -T) \propto
(\alpha^{-1})^{\delta}$.
The fitting curve for $\alpha \to \infty$ is shown in Figure~6, by the
dash-dotted curve,
with the estimated value $T_c(\alpha \rightarrow  \infty) =
0.155\Lambda$. The existence of the critical temperature, $T_c (\alpha \rightarrow
\infty)$, above which the chiral symmetry may always be restored, was also claimed
by Kondo and Yoshida [6].

\begin{table}
\caption{Critical exponents $\eta$ and $\nu$.}\label{tab_2}
\begin{center}
\begin{tabular}{cc|cc}
\hline
 $T/\Lambda$ & $\eta$ & $\alpha$ & $\nu$ \\
\hline \hline
  0.100 & 0.450 & 2.75 & 0.504 \\
  0.105 & 0.443 & 2.90 & 0.562 \\
  0.110 & 0.452 & 3.00 & 0.535 \\
  0.115 & 0.447 & 3.25 & 0.614 \\
  0.120 & 0.436 & 3.50 & 0.538 \\
  0.125 & 0.442 & 3.75 & 0.469 \\
  0.130 & 0.445 & 4.00 & 0.616 \\
  0.135 & 0.439 & 4.25 & 0.780 \\
  0.140 & 0.425 & 4.50 & 0.498 \\
  0.145 & 0.459 & 4.75 & 0.780 \\
  0.150 & 0.446 & 5.00 & 0.644 \\
\hline
\end{tabular}
\end{center}
\end{table}

The critical exponents are determined as shown in Table~II. As can be seen from
the table~II, the estemated value of the exponent $\eta$ is fairly stable over
the range of temperature $T$. The averaged value of the critical exponent $\eta$ is
\begin{equation}
 \eta = 0.44 \ .
\end{equation}
Above fact that the critical exponent $\eta$ is cleanly determined in the present
analysis, strongly supports the chiral phase transition of QED at finite temperature
under the change of the coupling constant to be of second order. As for the exponent
$\nu$, the estimated value is not so stable over the range of coupling constant
$\alpha$. The averaged value of the exponent $\nu$ is
\begin{equation}
\label{eq_15}
 \nu = 0.60 \ .
\end{equation}
We must improve our analysis to get stable values of $\nu$ in order to understand
clearly the nature of the temperature-dependent phase transition of QED.

\section{ Conclusion and discussion}

In this paper we investigated the consequences of the improved ladder (point vertex) DS
equation in QED for the fermion self-energy with the fully HTL resummed photon
propagator, where the gauge is fixed to the Landau gauge. We have made use of the
improved instantaneous exchanege (IE) approximation to the photon exchange, namely,
we have kept the exact HTL resummed transverse (magnetic) propagator, while taking
the IE limit to the logitudinal one. This approximation can account for the essential
temperature effect of the HTL resummation to the photon propagator. The prices
we should pay due to the two additional approximations, the ladder approximation and
the IE approximation, are the loss of the gauge invariance of the results, and the
limitation of the applicability of our analysis to the rather high temperature region.
Several discussion and comments on these points are given at the end of this section.

Before summarizing the results of the present investigation, we should stress the fact that the
existence of non-trivial imaginary part in the fermion self-energy function $\Sigma_R$ is essentially important in the analysis of dynamical (fermion) mass generation at finite temperature. As we have
shown in section II.3, we cannot naively neglect the imaginary part without facing the inconsistent
constraint equations.  

With this fact in mind, our results can be summarized as follows.

\begin{enumerate}
\item The chiral phase transition is found to proceed through the second order 
    transition. This can be seen with the generation of the dynamical fermion
    mass at the critical temperature or at the critical coupling constant
    without any discontinuity, see Figures 2 and 3. The clean determination of
    the critical exponents also strengthen this fact.
\item The critical temperature $T_c$ at fixed value of $\alpha$ is significantly
    lower than the previous results [6,7,8], namely, the restoration of chiral 
    symmetry occurs at lower temperature than previously expected. This
    fact shows that in the previous analyses the important temperature 
    effects were thrown away due to the inappropriate approximations.
\item The critical temperature $T_c$ increases as a function of the coupling 
    constant $\alpha$, with the saturation behavior in the stronger coupling
    constant region, see Figures 5 and 6. This fact may indicate that the 
    chiral symmetry is always restored at sufficiently high temperature, no 
    matter how large the coupling constant becomes, agreeing with the result
    of Ref.~[6] and contradicting that of Ref.~[8]. 
\item The critical curve determined, Figure 6, shows that the area of the chiral 
   broken (symmetric) phase is significantly smaller (larger) than 
   previously expected [6,7,8]. This fact proves that the effect of thermal 
   fluctuation to liberate the chiral symmetry is stronger than previously 
   considered.
\item The critical exponent $\eta$, that describes the phase transition
      at finite temperature under the change of the coupling constant,
      is determined cleanly, $\eta = 0.44$. This 
    fact also supports strongly the phase transition of QED at finite 
    temperature to be of second order.
\end{enumerate}

All the above results show the importance of correctly taking the thermal effects
into the analysis of chiral phase transition. Inclusion of only the HTL resummed
gauge boson (photon) propagator has significantly changed even the qualitative
behavior of the phase transition. This fact proves \textit{a posteriori} the
effectiveness of the DS equation in the HTL approximation, thus also indicates the
correctness of our research-strategy, namely, the importance of the full HTL resummed
DS equation analysis of the chiral phase transition at finite temperature/density.
Further investigation along the line of our strategy is needed to answer step by step
the question how we can closely take the essential thermal effects into the ``kernel''
of the DS equation.  

As noted above, however, in the present analysis, there are two shortcomings. One is
the loss of the gauge invariance of the result due to the ladder approximation, and
the other the limitation of the applicability of our analysis to the rather high
temperature region due to the IE approximation. The first improvement we should make
is to resolve these problems.

As shown in section III.3, see Figure~1 and Table~I, all the fermion wave function
renormalization constants receive more than 10-20 percent corrections. This fact means
$Z_2 \neq 1$, and shows the explicit violation of the gauge invariance of the result,
since $Z_1 =1$ due to the point-vertex approximation, thus indicating the necessity of the
gauge-parameter dependent analysis. In order to maximally respect the gauge invariance
of the result in the improved ladder approximation, we should keep explicitly the
gauge-parameter dependent contributions to the DS equations, Eqs.~(\ref{eq_8})-(\ref{eq_10}),
and solve them with the constraint $A(P) = 1$, namely, $Re[A(P)] = 1$ and $Im[A(P)] = 0$,
which guarantees $Z_2 = 1$ as required, being consistent with the Ward identity $Z_1=Z_2$.
This can be carried out by successively adjusting
the gauge-parameter $\xi$ in solving the above equations (, which may correspond to
choose some non-linear gauge). The result of this ``gauge invariant'' analysis within
the improved ladder approximation will be given in a separate paper [14].

The second problem of the limitation of the applicability of our analysis to the rather high
temperature region due to the IE approximation, casts the shadow on the reliability of
estimated value of $\alpha_c(T \rightarrow 0) = 2.576$. Actually, the $T \rightarrow 0$ limit
of the DS equations with the IE approximation, Eqs.~(\ref{eq_8})-(\ref{eq_10}),
do not coincide with the DS equations in the vacuum ($T=0$) theory. Thus to get more
reliable estimate of the critical coupling constant in the zero-temperature limit, we must
perform the analysis by getting rid of the IE approximation
to the photon propagation, which is now under invetigation.

\begin{center}
Acknowledgment
\end{center}

We thank the useful discussion at the Workshop on Thermal Quantum Filed
Theories and thier Applications, held at the Yukawa Institute for
Theoretical Physics, Kyoto, Japan, 8 -- 10 August 2002. This work is
partly supported by Grant-in-Aid of Nara University, 2002 (HN).
The numerical calculation is mainly done at the Computer Center, Nara University.


\begin{thebibliography}{99}
\bibitem{ref1} T. Maskawa and H. Nakajima, Prog. Theor. Phys. 52, 1326 (1974);
         54, 860 (1975);
         R. Fukuda and T. Kugo, Nucl. Phys. B117, 250 (1976).
\bibitem{ref2} K. Yamawaki, M. Bando and K. Matumoto, Phys. Rev. Lett. 56, 1335
        (1986);
         K.-I. Kondo, H. Mino and K. Yamawaki, Phys. Rev. D39, 2430 (1989).
\bibitem{ref3} W. A. Bardeen, C. N. Leung and S. T. Love, Phys. Rev. Lett., 56,
         1230 (1986);
         C. N. Leung, S. T. Love and W. A. Bardeen, Nucl. Phys. B273, 649 (1986);
         W. A. Bardeen, C. N. Leung and S. T. Love, Nucl. Phys. B323, 493 (1989).
\bibitem{ref4}A. Barducci, R. Casalbuoni, S. De Curtis, R. Gatto and G. Pettini,
         Phys. Rev. D41, 1610 (1990).
\bibitem{ref5} S. K. Kang, W.-H. Kye and J. K. Kim, Phys. Lett. B299, 358 (1993).
\bibitem{ref6} K.-I. Kondo and K. Yoshida, Int. J. Mod. Phys. A10, 199 (1995).
\bibitem{ref7} M. Harada and A. Shibata, Phys. Rev. D59, 014010 (1998).
\bibitem{ref8} K. Fukazawa, T. Inagaki, S. Mukaigawa and T. Muta, Prog. Theor. Phys.
         105, 979 (2001).
\bibitem{ref9} K.-C. Chou, Z.-B. Su, B.-L. Hao and L. Yu, Phys. Rep. 118, 1 (1985);
         L. V. Keldysh, Zh. Eksp. Teor. Fiz. 47, 1515 (1964) [Sov. Phys. JETP 20,
         1018 (1965)].
\bibitem{ref10} E. Braaten and R. D. Pisarski, Nucl. Phys. B337, 569 (1990); B339,
         310 (1990);
         J. Frenkel and J. C. Taylor, Nucl. Phys. B334, 199 (1990).
\bibitem{ref11} M. E. Carrington, and U. Heinz, Eur.Phys. J., C1,619 (1998); 
         Hou Defu, M. E. Carrington, R. Kobes, and U. Heinz, Phys. Rev. D61, 085013
        (2000).
\bibitem{ref12} Y. Fueki, H. Nakkagawa, H. Yokota and K. Yoshida, Prog. Theor. Phys.
        107, 759 (2002).
\bibitem{ref13} P. Aurenche, T. Becherrawy, E. Petitgirard, hep-ph/9403320
         (unpublished);
         A. Ayala and A. Bashir, Phys. Rev. D64 , 025015 (2001).
\bibitem{ref14} Y. Fueki, H. Nakkagawa, H. Yokota and K. Yoshida, to appear. 
\bibitem{ref15} Y. Fueki, H. Nakkagawa, H. Yokota and K. Yoshida, work in progress.
\bibitem{ref16} V. V. Klimov, Sov. J. Nucl.Phys. 33, 934 (1981); Sov. Phys. JETP 55,
         199 (1982);
         H. A. Weldon, Phys. Rev. D26, 1394 (1982); Phys. Rev. D26, 2789 (1982).
\bibitem{ref17} H. A. Weldon, Ann. Phys. (N.Y.) 271, 141 (1999).
\bibitem{ref18} Preliminary results were presented at the 4th Int'l Conference on
         Physics and Astrophysics of Quark-Gluon Plasmas (ICPAQGP-2001),
	held at Jaipur, India, November 2001,
         H. Nakkagawa, H. Yokota, K. Yoshida and  Y. Fueki, hep-ph/0112290, to appear
         in the proceedings.
\end{thebibliography}
\end{document}